# Detection of selective sweeps in cattle using genome-wide SNP data

**Holly R. Ramey[1], Jared E. Decker[1], Stephanie D. McKay[1,2], Megan M. Rolf[1,3], Robert D. Schnabel[1], Jeremy F. Taylor[1§]**

[1]Division of Animal Sciences, University of Missouri, Columbia MO 65211 USA

[2]Department of Animal Science, University of Vermont, Burlington, VT 05405 USA

[3]Department of Animal Science, Oklahoma State University, Stillwater, OK 74074 USA

[§]Corresponding author

Email addresses:

Holly R. Ramey (hrrxb2@missouri.edu)

Jared E. Decker (deckerje@missouri.edu)

Stephanie D. McKay (stephanie.mckay@uvm.edu)

Megan M. Rolf (mrolf@okstate.edu)

Robert D. Schnabel (schnabelr@missouri.edu)

Jeremy F. Taylor (taylorjerr@missouri.edu)

**Corresponding Author:**

Dr. Jeremy F. Taylor
S135B Animal Sciences
920 East Campus Drive
University of Missouri
Columbia MO 65211-5300
E-mail: taylorjerr@missouri.edu
Tel.: +1-573-884-4946
Fax: +1-573-882-6827

# Abstract

## Background

The domestication and subsequent selection by humans to create breeds and biological types of cattle undoubtedly altered the patterning of variation within their genomes. Strong selection to fix advantageous large-effect mutations underlying domesticability, breed characteristics or productivity created selective sweeps in which variation was lost in the chromosomal region flanking the selected allele. Selective sweeps have now been identified in the genomes of many animal species including humans, dogs, horses, and chickens. Here, we attempt to identify and characterize regions of the bovine genome that have been subjected to selective sweeps.

## Results

Two datasets were used for the discovery and validation of selective sweeps via the fixation of alleles at a series of contiguous SNP loci. BovineSNP50 data were used to identify 28 putative sweep regions among 14 diverse cattle breeds. To validate these regions, Affymetrix BOS 1 prescreening assay data for five breeds were used to identify 114 regions and validate 5 regions identified using the BovineSNP50 data. Many genes are located within these regions and the lack of sequence data for the analyzed breeds precludes the nomination of selected genes or variants and limits the prediction of the selected phenotypes. However, phenotypes that we predict to have historically been under strong selection include horned-polled, coat color, stature, ear morphology, and behavior.

## Conclusions

The identified selective sweeps represent recent events associated with breed formation rather than ancient events associated with domestication. No sweep regions were shared between indicine and taurine breeds reflecting their divergent selection histories and environmental habitats to which these sub-species have adapted. While some sweep regions have previously been identified using different methods, a primary finding of this study is the sensitivity of results to assay resolution. Despite the bias towards common SNPs in the BovineSNP50 design, false positive sweep regions appear to be common due to the limited resolution of the assay. Furthermore, this assay design bias leads to the majority of detected sweep regions being breed specific, or limited to only a small number of breeds, restricting the suite of selected phenotypes that can be detected to primarily those associated with breed characteristics.

## Background

The transition from hunter-gather lifestyles to permanent dwelling societies was facilitated by both plant and animal domestication [1]. The domestication of cattle occurred between 8,000 and 10,000 years ago and led to changes in the genome of the species due to the effects of demography and selection [2, 3]. Much of the variation within the genetically diverse ancestral population was either lost due to the limited sampling of animals within the sites of domestication or was partitioned into the subpopulations which went on to become recognized as distinct breeds. Selection for the phenotypes contributing to domesticability, biological type (draft, milk, meat) and the aesthetically appealing morphologies that have become breed hallmarks (polled, coat color and patterning [4-8]) have also impacted the extent and distribution of variability within the genome.

Strong on-going selection for variants of large effect leads to a loss of variation within the chromosomal region flanking the selected variant and ultimately the complete fixation of the haplotype which harbors the variant. This phenomenon is known as the “hitch-hiking effect” [9] and the regions of the genome in which selection has driven a haplotype to complete fixation are defined as having been subjected to “selective sweeps.” Studies of selective sweeps differ from the classical forward genetic approach which progresses from a phenotype to the identification of the underlying causal genes or mutations. Rather, they follow a reverse genetics approach that begins with the signature of selection and attempts to infer the selected mutation and its associated phenotype [1].

Several approaches have been used to identify regions of the genome which have been subjected to selective sweeps, including those based on allele frequency spectra, linkage disequilibrium, haplotype structure, and employing sophisticated models fit using, for example, maximum composite likelihood estimation [10-12]. These approaches require the use of high-density single nucleotide polymorphism (SNP) data which has previously been shown to be useful for detecting selective sweeps in human populations [13, 14]. Studies aimed at localizing selective sweeps have been performed in many animal species using SNP and microsatellite loci. In chicken, selective sweeps have been found to involve loci believed to be involved in domestication and include *BCDO2* which controls yellow and white skin colors, *SEMA3A* which plays a role in axonal pathfinding important in brain development, and *THSR* which is postulated to derestrict the regulation of seasonal reproduction [15]. Selective sweeps have been found in the dog genome at *TRYP1* which controls black coat color in Large Munsterlanders and in *FGFR3* in Daschshunds [16]. *FGFR3* mutations cause achondroplasia in humans and cattle. Other studies in dogs have identified a sweep surrounding *IGF1* which is responsible for size variation [17] and in a region for which the selected phenotype is unknown in Boxers [18]. These sweeps range in size from 28 kb to 40 Mb suggesting considerable variation in the intensity of selection and also in the population census size. A ~75 kb selective sweep at a locus influencing stature has also been found in the horse genome in a region upstream of a transcription factor (*LCORL*) that is associated with variation in human height [19]. A 28 kb selective sweep in a region of the swine genome harboring *IGF2* has also been found for which the underlying selected phenotypes are increased muscle mass and decreased fat deposition [20]. Domestic animals have been demonstrated to be excellent models for genetic studies due to the availability of extensive pedigrees and because, as species, they are frequently more genetically diverse than humans [21]. However, relatively few selective sweep studies have been conducted to date in livestock species to elucidate the genes which have been selected by humans to create the diverse spectrum of breeds and specialized biological types for draft, milk and meat production.

In this study, we sought to identify the signatures of selective sweeps in cattle as genomic regions in which contiguous SNPs from the BovineSNP50 assay were all individually fixed for a single allele within a sample of 6,373 genotyped individuals representing 14 breeds (Table 1). Our goals are ultimately to identify the selected mutations and phenotypes which were selected by our ancestral herdsmen in the processes of domestication and formation of breeds and biological types.

# Results

## Regions identified as harboring selective sweeps using BovineSNP50 data

Twenty eight genomic regions on 15 chromosomes were identified as putatively harboring selective sweeps (Table 2). Selective sweeps were found in all 14 breeds; however, breed-specific selective sweeps were not identified in every breed. Twenty three predicted sweeps were breed-specific and 5 were shared among two to seven breeds with 4 of those containing three or more (Figure 1). Breed-specific sweeps averaged 336,263 bp and ranged in size from 207,050 to 702,424 bp whereas sweeps common to two or more breeds averaged 441,314 bp and ranged in size from 214,554 to 866,260 bp. The haplotypes found at the core loci involved in each of these putative selective sweeps were identical for each of the breeds in which the sweep was detected.

Three of the five selective sweep regions detected in two or more breeds involved both beef and dairy breeds, whereas the 358 kb region on BTA12 is common to only the Angus, Salers, Shorthorn and Simmental beef breeds. None of the five selective sweeps shared by two or more breeds are phylogenetically congruent in the sense that we might have expected the sweep to have arisen in a recent common ancestor [22]. While the large selective sweep region on chromosome 6 at ~75.9-76.7 Mb is shared by the closely related Salers and Brown Swiss breeds, Salers and Limousin are sister breeds [2, 3] and Limousin does not demonstrate evidence of this sweep. There were no putative sweeps shared in common between any of the taurine breeds with the indicine Brahman breed. The DAVID functional analysis did not yield any significant functional enrichment of gene ontology terms for these sweep regions.

**Validation of selective sweep regions using Affymetrix prescreening assay data**

To validate the selective sweep regions identified using the BovineSNP50 data, we examined ultra high-density SNP genotype data produced by the Affymetrix prescreening assay used for the design of the BOS 1 array (AFFXB1P) for small samples from five breeds, of which four were included in our BovineSNP50 data (Table 1). Since only a subset of the 14 breeds genotyped with the BovineSNP50 assay were also assayed with the AFFXB1P assay, we had the potential to validate only 11 of the putative selective sweep regions identified in Table 2 and five regions were confirmed (Table 3). Only two of the regions were confirmed in the same breeds that led to their identification using the BovineSNP50 assay and for two of the remaining regions, discovery occurred in Hanwoo and confirmation occurred in the phylogenetically similar Wagyu breed [2, 3]. However, the region on BTA13 was identified in Holstein using the BovineSNP50 data but was independently validated in Wagyu by the AFFXB1P data.

**Selective sweep regions independently identified using the AFFXB1P data**

A total of 86 putative selective sweep regions spanning from 200,088 to 846,420 bp and averaging 321,001 bp were identified on 28 of the 29 bovine autosomes in the genotyped 5 breeds and, of these regions, 20 were shared in two or more breeds (Tables 4 and 5). These regions harbored from 20 to 477 contiguous SNPs with no more than 5% of the SNPs being variable. Among the selective sweeps identified in two or more breeds, the number of breeds included in this analysis was too small to make inferences about the phylogenetic congruence of shared sweep regions, however, three sweeps were found only in the closely related East Asian Wagyu and Hanwoo breeds. All of the breeds that shared a common selective sweep (Table 5) were fixed for the same core haplotype with two exceptions. The sweep on BTA16 at 45,386,065-45,652,672 and on BTA21 at 1,727,412-2,142,823, both shared by the Angus and Simmental breeds, were fixed for haplotypes that differed at a single SNP. The allele that varied in the haplotypes that were swept to fixation on BTA16 was the 7th of 248 SNPs, whereas on BTA21 the variable allele was at the 22nd of 30 SNPs, indicating a conserved core at both loci. No sweeps were found in common between the four taurine breeds and the indicine Brahman.

## Discussion

The BovineSNP50 data identified a sweep region towards the centromere of BTA1 harboring 11 contiguous monomorphic SNPs and spanning 301 kb in Angus (Figure 2). This region contains the *polled* locus [6] for which this breed has been strongly selected for homozygosity of the *polled* allele [23]. Hereford cattle are also homozygous for the dominant *spotted* allele at the *spotted* locus which is a candidate for the 210 kb sweep region at 70.65-70.87 Mb on BTA6 [8]. The *spotted* locus affects the white points on the face, underline, feet and tail which are a definitive characteristic of the Hereford breed. These breed-specific sweeps are clearly examples of strong selection on loci which underlie phenotypes that are hallmarks of certain breeds and where the underlying causal mutation is known or has been mapped to a chromosomal location.

Several regions in which we detected putative sweeps using the BovineSNP50 data contain no annotated genes and either this reflects the incomplete annotation of the bovine genome, or the fact that the selected functional mutation within each of these regions is not located within a protein coding gene. Examining each of these regions for the alignment of human, pig and sheep mRNA orthologs failed to identify any. Recent work has identified ncRNAs which regulate the expression of nearby genes [24] and may help identify candidates for the mutations in these regions that were subjected to selection. The putative sweep shared among six breeds on BTA24 contains four pseudogenes, of which three are olfactory receptor-like. Whether any of these pseudogenes are expressed is unclear; however, olfactory receptor loci were detected as being recently duplicated within the bovine genome [25] suggesting that they may also be under strong selection for newly evolving functions. The common sweep region on BTA12 contains neurobeachin (*NBEA*) and mab21-like 1 (*MAB21L1*) which have been implicated in human autism and psychiatric disorders, respectively [26-28]. Since these phenotypes represent extreme behaviors, it is intriguing to speculate that mutations in these genes may also predispose cattle to increased docility and more favorable temperaments when handled by man.

A selective sweep on BTA13 was detected in Holstein using the BovineSNP50 data and was also discovered in Wagyu using the AFFXB1P data. The region from 15.49-15.74 Mb contains diacylglycerol kinase zeta (*DGKZ*) and several bovine ESTs. *DGKZ* has been implicated as a member of the downstream leptin signaling pathway and reduced expression or activity within the hypothalamus has been associated with obesity [29]. The Holstein and Wagyu breeds are phylogenetically distant, however, Wagyu are believed to have been influenced by several European taurine breeds, including Holstein, during the late 1800s and both breeds are known for their ability to store intramuscular fat without accumulating excessive subcutaneous fat. The 351 kb selective sweep region from 1.67-2.02 Mb on BTA1 found in Angus using the BovineSNP50 and validated in Angus using the AFFXB1P data contains a fixed 321 marker haplotype which harbors the *polled* locus [6] a hallmark of the breed which contains only polled animals.

A region on BTA18 from 14.72-14.97 Mb was detected to harbor a selective sweep in Hanwoo cattle using the BovineSNP50 data and rediscovered in Angus and Simmental, using the AFFXB1P data. This 248 kb region contains several annotated genes (Table 3), but importantly harbors melanocortin 1 receptor (*MC1R*) in which mutations lead to the black coat color in cattle [7]. Angus have been strongly selected for black coat color and almost all registered animals are now homozygous black confirming that the basis for this selective sweep in Angus was for the black coat color allele. The American Simmental Association registers crossbred animals and many breeders have developed Simmental × Angus crossbreds to capitalize on the premium that carcasses from black coated cattle can achieve if they qualify for Angus branded products. While all of the Simmentals genotyped with the BovineSNP50 assay were purebred, of the 6 Simmental animals genotyped with the AFFXB1P assay, one was homozygous black, four were heterozygous and one was homozygous red. Furthermore, the Hanwoo animals are all yellow coated. The fact that the sweep in Angus was not found using the BovineSNP50 data suggests a resolution issue with the requirement that at least 6 contiguous loci spanning at least 200 kb be fixed in order to declare a sweep. On the other hand, the fact that a sweep was detected in the AFFXB1P data for Simmental that was not detected using the BovineSNP50 data suggests a sampling issue since of the 12 chromosomes representing this breed, only 6 originated in Simmental for this region of the genome. The result for Hanwoo is more interesting since the sweep was declared in Hanwoo using the BovineSNP50 data for 48 individuals but was not confirmed in the 11 individuals genotyped with the AFFXB1P data. This suggests that either the region is not correctly assembled, or that an ancient breed foundation event may have occurred in which the yellow allele was fixed in this breed, but that sufficient mutation events have occurred on this *MC1R* haplotype to cause it to fail to be detected as a sweep using the high density data. Finally, of particular interest is the fact that no sweep was identified in this genomic region in Wagyu cattle suggesting that black coat color in Angus and Wagyu cattle may not be allelic. Recently, a mutation within β-defensin 103 (*CBD103*) has been shown to cause black coat color in dogs [30]. The cattle ortholog of *CBD103* maps to 4.89 Mb on BTA27 centromeric of the sweep that was detected in Wagyu cattle (Table 4).

A region on BTA16 at 45.39-45.65 Mb was identified as containing a sweep in six breeds including Angus and Simmental in the BovineSNP50 data and was validated in Angus and Simmental using the AFFXB1P data. This region contains *RERE* a member of the atrophin family of arginine-glutamic acid dipeptide repeat-containing proteins that is implicated in embryogenesis and embryonic survival in drosophila, mice and zebrafish [31-33]. *RERE* encodes a tissue-specific transcriptional corepressor, which in association with histone deacetylase, directs growth factor activity and early embryonic patterning [31, 33]. A second sweep was also predicted nearby on BTA16 in Hanwoo and identified at 52.63-52.86 Mb using the AFFXB1P data in Simmental and Wagyu, but not in Hanwoo. However, this 228 kb region includes only 30 SNPs in the AFFXB1P assay, a relatively small number of SNPs compared to the 107-321 SNPs in the other four sweep regions validated using the AFFXB1P data. The region contains 11 protein coding genes including agrin (*AGRN*), *ISG15* ubiquitin-like modifier, and hairy and enhancer of split 4 in *Drosophila* (*HES4*). Two of these genes are involved in embryonic development; *AGRN* functions in the process of cell adhesion between the embryo and maternal tissues [34] while *HES4* is a member of a family of genes known to be oscillators and repressors that affect embryonic developmental events [35]. The *ISG15* ubiquitin-like modifier encodes a protein that is present in the uterus of many species, including ruminants, during early pregnancy and is one of the many proteins responsible for acknowledging and maintaining pregnancy [36, 37]. The proximity of the two BTA16 sweep regions and involvement of their constituent genes in embryonic development may not be coincidental.

For the 66 breed-specific regions identified within the AFFXB1P data (Table 5), many regions contained several genes while a few contained no annotated genes. Five regions contained only one annotated gene or pseudogene as well as some miscellaneous RNAs and uncharacterized protein coding regions. Of these, three sweep regions predicted in Wagyu on BTA1 and BTA27 contained pseudogenes. The sweep region predicted in Wagyu on BTA27 contains a pseudogene of chromobox homolog (*CBX1*) which is a gene involved in the epigenetic reprogramming of the genome during embryo development and later regulates gene expression [38-40]. One sweep region on BTA3 predicted in Angus contains *TSPAN2*. Tetraspanin-2 is involved in oligodendrocyte signaling and maturation [41, 42]. The fifth predicted sweep region harboring only a single gene was found on BTA28 in Simmental. This region contains adenosine kinase (*ADK*) which catalyzes phosphate transfer from ATP to adenosine and controls the concentrations of adenine nucleotides both intra- and extra-cellularly. Variation in adenosine abundance affects many physiological processes such as rate of blood flow throughout the body and neurotransmission within the brain [43]. Probably the most interesting gene within this group is found on BTA11in Wagyu. Neurexin 1 (*NRXN1*) is located in the sweep; mutations in neurexin genes have been theoretically linked to cognitive diseases since they are important in synaptic function [44]. Mutations in human *NRXN1* have been confirmed to cause autism spectrum disorders via neurotransmission activity [45]. The link between genes involved in psychiatric disorders poses a potential link to selection based on cattle temperament. Mutations in these genes may confer behavioral changes that are conducive to improved temperament when handled by humans and could have been selected during domestication or more recently following breed formation to develop more manageable animals.

When human and porcine orthologs were sought in regions found to contain a single annotated bovine gene, only one region was found to harbor additional loci in human. The sweep region on BTA1 is orthologous to a region on the long arm of HSA9 which contains the iron-sulfur cluster assembly 1 homolog (*ISCA1*). This gene is important for functionality of a mitochondrial protein in the Fe-S pathway and mutations within *ISCA1* cause many conditions and diseases in humans [46]. All of the remaining regions appear to contain only a single gene and these genes are strong candidates for harboring mutations which cause large phenotypic effects that were subjected to strong selective sweeps in these breeds. However, it is also possible that the selected variants are not actually located within coding regions of the genome and are, for example, within non-coding RNAs associated with regulatory genes [24].

Among the other detected breed-specific sweep regions, a region on BTA14 specific to Angus cattle harbors several genes including *PLAG1* which has been associated with variation in human height [47] and the stature of cattle [48]. Other positional candidates for the stature QTL in this region include *RSP20* and *SDR16C5* [48]. Of all of the US cattle breeds, Angus has been the most strongly selected for growth and frame size during the last 30 years [49] perhaps creating this selective sweep. The telomeric sweep region on BTA1 found in Angus contains several genes, including *LEKR1*and *CCNL1* in which mutations have been associated with reduced birth weight in humans [50]. Angus cattle have recently been selected to reduce birth weights to ease dystocia and difficulties in birthing and raising calves which are heavier at birth [22]. In Simmental cattle, we found a sweep region on BTA25 that is orthologous to a region on HSA16 which contains genes involved in brain function and neurodevelopment [51-53] that have been associated with autism spectrum disorders. This provides further support for the speculation that cattle have been selected to improve their temperament and docility if we are prepared to accept that mutations in these genes may have less severe behavioral consequences than those associated with autism disorders. Finally, we found three regions harboring selective sweeps in Brahman. A region on BTA10 contains four predicted protein coding domains, none of which are annotated. A second region on BTA22 at 10.70-10.96 Mb contains four genes: *TRANK1, DCLK3, GOLGA4*, and *ITGA9. TRANK1* lies within a region in humans which contains a SNP that is significantly associated with bipolar disorder [54] while doublecortin-like kinase 3 is expressed throughout brain development and in the adult brain [55]. Golgin-4 is the golgin family member that is implicated in providing structure via tethering within the Golgi [56] and integrin alpha 9 is implicated in the reproductive process of gamete interaction. *ITGA9* is expressed on the surface of mouse and human eggs and reduced expression decreases the binding ability of the sperm which reduces fertilization success [57]. Once again, the genes located in this region appear to be potentially associated with behavior or reproductive success, a key fitness trait. Perhaps the most interesting of the three selective sweep regions detected in Brahman is located at 48.68-48.90 Mb on BTA5 and contains the methionine sulfoxide reductase B3 (*MSRB3*) gene which has previously been identified as a candidate for a QTL affecting ear floppiness and morphology in dogs [58, 59]. Brahman cattle were developed in the US as a cross between the *Bos taurus indicus* breeds Guzerat, Nellore, Gir and Indu Brazil imported primarily from Brazil but all originating in India and the *Bos taurus taurus* Shorthorn and Hereford breeds [60]. There is considerable variation among these breeds for ear length and morphology with Indu Brazil animals having particularly large, pendulous ears. Thus, the sweep in this region may reflect strong recent selection by breeders to establish a specific Brahman ear morphology type.

Of the four regions detected in three or more breeds in the AFFXB1P data, the first on BTA6 was located at 5.64-5.99 Mb which was shared by all four genotyped taurine breeds. It harbors what appears to be a large intron belonging to *MAD2L1* which is required for the proper onset of anaphase during cell division. Within the second region on BTA6 at 106.84-107.31 Mb *RGS12*, a member of the regulator of G protein signaling gene family that directs many cellular activities including hormone signaling, is a positional candidate for an ovulation rate QTL in swine [61, 62]. The detected sweep with a common core at 0.21-0.45 Mb on BTA26 shared by all four taurine breeds contains a series of olfactory receptor loci. The sweep region with a common core at 0.32-0.56 Mb on BTA29 detected in Simmental, Wagyu, and Hanwoo and contains only ncRNA genes ankyrin repeat domain 26 pseudogene 3 (*ANKRD26P3*) and coiled-coil domain containing 144C (*CCDC144C*) towards the centromeric end of the region. However towards the center of the predicted sweep region is a region that is highly conserved between human and cow and may represent a long ncRNA. The DAVID functional analysis found no enrichment of ontological terms for the genes located within these selective sweep regions that are shared among multiple breeds and this was expected assuming that each selective sweep was based upon different phenotypes.

The presence of olfactory pseudogenes within sweep regions is intriguing and suggests that olfactory loci play a major role in the domesticability of species. Olfactory receptor (OR) genes have previously been found to have been under selection in cattle [12, 25] and more recently in swine [63] and it has been hypothesized that pigs rely intensely on their sense of smell for scavenging. Alterations in the need for wild animals to search for food after their domestication may result in a relaxation of the need for purifying selection acting on these genes allowing them to freely evolve in odorant and tastant detection or lack thereof. In tetrapods, anywhere from 20 to 50% of OR loci exist as pseudogenes [64] and while it is not clear if these genes were ever functional, the acquisition of trichromatic vision has been postulated as facilitating the loss of OR genes [65]. On the other hand, cattle are dichromatic and yet still have a high proportion of OR pseudogenes [64] some of which were found to have been under selection in this study. If these pseudogenes lack functionality, we might have expected them to have been deleted from the genome or to have been significantly disrupted by mutation. However, as many as 67% of OR pseudogenes are expressed in human olfactory epithelium [66] suggesting that similar percentages of bovine OR pseudogenes are also expressed and that many of these loci are functional and rapidly evolving in copy number [25].

We utilized two genotyping assays to identify putative selective sweep regions within the bovine genome. The BovineSNP50 assay was employed because we have genotyped a large number of registered animals from several breeds with this assay; however, we recognize that the assay is not ideal for this purpose due to the ascertainment of common SNPs in its design. Since the *Bos taurus taurus* breeds in Table 1, and Angus and Holstein in particular, were used for SNP discovery and SNPs with high minor allele frequencies in these breeds were preferentially included during the design of the assay [14, 15] it is clearly unsuited to the identification of selective sweep regions that might be common among breeds. However, SNPs were included in the design of the BovineSNP50 assay if they were found to be variable in several, but not necessarily all of these breeds. Therefore, the assay theoretically possesses the ability to identify selective sweeps that are specific to individual breeds or to a small number of breeds. However, rather than characterizing sweeps that occurred during the domestication of cattle and that should therefore be common, e.g., among European taurine breeds that descended from cattle that were domesticated in the Fertile Crescent, these sweeps are much more likely to have occurred during the formation of breeds and will reflect selection to fix phenotypes such as coat color or the absence of horns within specific breeds. A second limitation of this assay is that of calibration relative to the size of the sweep regions. While strong sweeps in numerically small populations are expected to result in the fixation of large haplotypes, weak selection in numerically large populations will result in the fixation of only a small core haplotype which may not be detected using the assay. Thus historic variation in the census population size among breeds may have resulted in variation in the size of the fixed haplotype and the inability to detect small haplotypes. By requiring N contiguous loci to each have a minor allele frequency (MAF) < □, for small □, we must choose N to be sufficiently large that it would be highly unlikely to observe N contiguous loci all with a MAF < □ due to chance alone and yet sufficiently small that the targeted sweeps are not smaller than 37 × (N-1) kb, where 37 kb represents the median intermarker interval on the BovineSNP50 assay. The design of the BovineSNP50 assay also led to lower average MAF and larger numbers of monomorphic loci in breeds such as Brahman, that are phylogenetically distant from the SNP discovery breeds [14]. To adjust for this bias, we defined N separately for each breed (Table 1) requiring larger N for breeds with larger numbers of monomorphic and low MAF loci. The definition of □> 0 is also important to this discussion since in the detection of sweeps we must allow for old sweeps in which *de novo* mutations may have begun to accumulate on the fixed haplotype, genotyping errors which are locus specific but average about 0.5% for this assay, and the incorrect ordering of loci by the UMD3.1 sequence assembly. Errors in the assembly are vastly more likely to cause false negative than false positive sweeps by incorrectly introducing a variable locus into a region of dramatically reduced variability within the genome.

We also employed the Affymetrix BOS 1 prescreening assay which contained almost 2.8 million putative SNPs that were screened for variability in a small number of animals from several breeds prior to the design of the commercial BOS 1 Axiom assay. While we had many fewer animals genotyped with this assay which influenced the estimation of MAF, the AFFXB1P assay had over 50× the number of SNPs present on the BovineSNP50 assay which offered considerably greater power for identifying small sweeps and the application of this assay also suffers less from ascertainment bias. While loci that have been fixed in all domesticated cattle relative to their auroch forbears will still not appear on this assay due to the requirement that the putative SNP must have been predicted to have been variable in the sequence data for at least one breed, there was much less selection for SNPs with high MAF in numerous breeds in the design of this assay relative to the BovineSNP50. Consequently, we expected this assay to identify putative sweep regions that could not be identified by the application of the BovineSNP50 assay and to more precisely define the boundaries of sweeps that were detected by the BovineSNP50 assay and validated by the AFFXB1P assay.

We found evidence for selective sweeps in genomic regions that were detected to have diverged between breeds using integrated haplotype scores (iHS) and $F_{ST}$ statistics in the Bovine HapMap project [12]. Putative sweep regions overlapped on chromosomes 2, 11, 12, and 14 detected either by an extreme $F_{ST}$ or iHS value. We found evidence for only one selective sweep region for which the Bovine HapMap project [11] identified divergence between breeds using $F_{ST}$ statistics. Using the BovineSNP50 data, we found a putative sweep in Angus, Salers, Shorthorn and Simmental cattle on BTA12 in a region harboring *NBEA* that was found to have differentiated among similar taurine breeds. The fact that only one such region was detected probably reflects the fact that our approach requires a completed sweep to have occurred whereas use of the $F_{ST}$ statistic can detect an ongoing sweep for which breeds have diverged.

No putative selective sweep regions were found in common between Brahman and any of the *B. t. taurus* breeds which likely reflects the recent admixture that occurred in the formation of the Brahman and the fact that the breed does not share a common phenotype such as coat color with any of the taurine breeds. Furthermore, indicine cattle are more commonly found in the southern tier of the US where they are exposed to higher temperatures and humidities and lower pasture qualities and availability than are taurine cattle which are more frequently found in the northern US. Consequently, we would not expect these breeds to have been subjected to selection for common morphological or adaptive phenotypes. Additionally, no common sweeps were detected between the cattle sub-species possibly due to the more severe ascertainment bias on MAF for BovineSNP50 loci in Brahman cattle. While SNP discovery was performed in Brahman for the development of the AFFXB1P assay, the number of indicine breeds sequenced for SNP discovery was small relative to the number of sequenced taurine breeds leading to a bias towards SNPs common in taurine cattle being included on the assay. However, the density of SNPs on this assay is so great (~1 SNP / kb) that we did not expect the reliability of sweep regions identified in Brahman to be significantly less than those identified in taurine breeds. Our identification of a putative sweep region harboring a previously identified QTL for ear length and floppiness in Brahman is consistent with the introduction of the undesirable allele from Indu Brazil cattle during breed formation and subsequent strong selection by breeders to remove the allele and fix a shorter ear type within the breed.

Low density assays, such as the BovineSNP50 assay, have previously been utilized to identify runs of homozygosity (ROH) within cattle breeds. The BovineSNP50 assay was found to be adequate for finding ROH and estimating inbreeding coefficients in cattle [67]. However, this is not the case for the detection of selective sweeps which typically span smaller regions of the genome than ROH which are frequently due to consanguinity and which may represent as much as 12.39% of the genome [67]. The use of low density SNP data for the detection of selective sweeps appears to be inadequate primarily due to the poor calibration of SNP density relative to the size of the targeted sweep regions such that any relaxation of the number of contiguous SNPs with fixed alleles required to detect smaller sweep regions leads to an elevated type I error rate. Strong, recent selective sweeps causing the fixation of large haplotypes may be identified using low-density SNP panels (~50,000 SNPs), however, older sweeps which have accumulated new mutations and weak sweeps which have resulted in the fixation of relatively small haplotypes will be missed.

Several putative selective sweeps identified using the BovineSNP50 assay failed to be validated using the AFFXB1P assay. While many fewer animals representing each breed were genotyped using the AFFXB1P assay, among the 50× additional SNPs within each such region, we found that at least 5% of the SNP had a MAF > $(2M)^{-1}$ where M is the number of genotyped animals. We have previously found that the genotyping error rate of loci on the Affymetrix Axiom BOS 1 assay is very similar to that of the Illumina BovineSNP50 and HD assays (~0.5%, data not shown) and thus, we do not expect genotyping errors to explain this result, although it is certainly a possibility. It appears that the phenomenon is random and associated with the number of regions expected to meet the criteria for declaring a sweep when ~8,000 of the 54,001 BovineSNP50 loci are monomorphic within a breed (Table 1) [68, 69]. For example, if 15% of SNPs are monomorphic within a breed, the probability that N contiguous SNPs are monomorphic is $0.15^N$, assuming independence, and in testing 50,000 SNP on 29 autosomes we would expect to find $0.15^N \times (50{,}000 - 29 \times (N - 1))$ regions in which N contiguous SNPs had fixed alleles. For N = 5 this corresponds to 4 false positives per breed and 0.6 false positives per breed when N = 6 (Table 1). As a consequence, the reliability of declaration of a selective sweep is dramatically improved when sweeps are found to be common between breeds, particularly when the breeds are phylogenetically distant. We found several sweep regions that were common to two or more breeds and five predicted sweeps detected by the BovineSNP50 assay were validated by the AFFXB1P data.

Identifying the mutations that underlie these sweep regions will be paramount to more fully understanding the effects of human interaction on the genomes of domesticated cattle. Candidates will soon become available by sequencing the genomes of individuals that are homozygous for identical SNP haplotypes within a sweep region but where some originate from the breeds predicted to have undergone a selective sweep and the others from breeds in which no sweep was detected. We found many regions to contain pseudogenes and protein coding regions but others that contained no annotated genes. While this reflects the nascent state of annotation of the bovine genome, it also reflects the fact that the mutations we seek may not be protein coding. For example, the mutation within the sweep region on BTA1 in Angus cattle that removes horns from these animals appears to be a ncRNA [6]. However, even after these mutations have been identified, our understanding of the phenotype that was created and selected to complete fixation may still be limited. The functional analyses of the genes within the selected regions sheds little light on this since each region likely harbored variants which influence different phenotypes. Finally, while our sampling of breeds was small, we found little evidence for the sharing of sweeps among phylogenetically closely related breeds. This suggests that the sweeps are mostly recent, post-dating breed formation and occurred for a variety of phenotypes for which selective value differed among the humans who developed these breeds.

## Conclusions

We identified selective sweeps that appear to primarily have been due to breed formation events. Due to the constraint that SNPs be variable in multiple breeds imposed during the design of the utilized assays, we did not identify any sweeps that were common to all breeds within the study. There were also no sweep regions predicted to be in common between breeds of taurine and indicine descent probably reflecting the different environmental and demographic forces to which these sub-species have been exposed during breed formation. For several of the detected regions we are able to identify the phenotypes and genes that were subjected to selection or to propose these based upon the results of previous mapping studies. However, for many of these regions the selected gene and phenotype are unclear. The fact that so many of the detected sweep regions harbor genes associated with reproductive processes or embryonic development is probably not remarkable considering the fact that strong selection acts on these fitness traits and that the time required to achieve fixation of variants of modest effect may be considerably longer than the 200 years since breed formation during which strong human selection has acted.

We demonstrate that the resolution and SNP ascertainment bias inherent in the design of the assay used to detect selective sweeps is of paramount importance and that the BovineSNP50 assay is not generally suitable for this purpose due to the high type I error rates that are likely to be encountered. SNP ascertainment bias leads to lower MAF in breeds that are phylogenetically distant from the SNP discovery breeds and an increased rate of monomorphic SNPs with these breeds. As whole genome sequencing becomes less expensive these problems will likely be ameliorated by sequencing a few distantly related individuals from each breed and this approach may also be used to identify candidate mutations underlying each sweep. However, the approach is reliant on the alignment of sequences to a Hereford reference assembly [25] which introduces a new set of biases unless *de novo* sequence assemblies can accurately be created for each breed.

The identification of genes and variants underlying historical selective sweeps is of interest from the perspective of understanding how human interaction with cattle has influenced the patterning of variation within the bovine genome. Perhaps of more importance, the discovery of the selected variants will lead to the identification of large effect QTLs and ultimately a better understanding as to the phenotypes which are affected by variation within genes and regulatory elements.

## Materials and Methods

### Samples, design, and genotyping

We utilized two data sets comprising high-density SNPs scored in animals that were registered by their respective breed societies. The first data set comprised 6,373 animals from 13 taurine breeds including Angus, Braunvieh, Charolais, Hanwoo, Hereford, Limousin, Salers, Shorthorn, Simmental, Brown Swiss, Finnish Ayrshire, Holstein, and Jersey, and the indicine Brahman breed (Table 1). Genotypes scored in these animals were generated using the Illumina BovineSNP50 BeadChip which assayed 54,001 loci with a median intermarker interval of 37 kb [68, 69]. The second data set which was used both for the discovery of putative selective sweep regions and for validation of results obtained from the analysis of the BovineSNP50 data set comprised 58 animals from the Angus, Hanwoo, Simmental, Wagyu, and Brahman breeds (Table 2) which were genotyped with the Affymetrix prescreening assay comprising 2,787,037 SNPs with a median intermarker interval of 975 bp that was used to generate allele frequency estimates for the design of the Axiom Genome-Wide BOS 1 assay [70].

The sampled breeds were chosen based on their geographical origins, historical uses by human, diverse phylogenetic relationships and because of the availability of at least 40 BovineSNP50 genotyped individuals. Each BovineSNP50 genotyped individual was registered with its respective breed association and was proven by pedigree-analysis to be purebred, since some associations (e.g., Simmental, Limousin) allow the registration of crossbred cattle. This sampling strategy was employed to ensure that there would be minimal effects of recent introgression between the breeds following breed formation.

**SNP filtering**

All X-linked loci were removed from the analysis due to the greater number of assembly issues that are associated with this chromosome and also because the studied animals were male resulting in a halving of the number of chromosomes sampled for each breed which leads to a reduction in the precision of allele frequency estimation. The remaining BovineSNP50 genotypes were filtered on call rate < 85% which left a total of 52,942 SNPs. Within the higher density

AFFXB1P data, we required SNPs to have a call rate of ≥85% across all 5 breeds and a minimum call rate of 50% within each of the individual breeds. Following filtering 2,575,339 SNPs remained.

## Identification of putative selective sweep regions

The BovineSNP50 data were analyzed by breed to identify putative selective sweeps. Because the number of variable loci differed within each breed primarily due to the breed of origin of SNP discovery in the design of the assay [14, 15], we required a breed-specific number (Table 1), and at a minimum 5, contiguous SNPs spanning at least 200 kb based upon UMD3.1 coordinates for which no SNP had a minor allele frequency MAF > 0.01 to declare a selective sweep. Following the breed-specific analysis, an across-breed comparison was performed in which any putative selective sweep which overlapped in two or more of the breeds was considered a common sweep and the selected haplotypes present within each breed were identified.

## Validation of selective sweeps

To validate the putative sweep regions detected using the BovineSNP50 data, we used a high-density data set representing SNPs genotyped at approximately a 50-fold increase in resolution, but scored in only a small subsample from 5 breeds. However, the analysis performed with these data was not guided by the results of the analysis of the BovineSNP50 data. We independently analyzed the SNPs as conducted for the BovineSNP50 data to identify putative selective sweeps and then compared the results of the analysis of the AFFXB1P data to those from the analyses of the BovineSNP50 data to identify putative selective sweeps identified within each breed by both analyses. To declare a putative selective sweep region we required the region to harbor at least 20 contiguous SNPs spanning at least 100 kb, with no more than 5% of the SNP having a MAF >

$(2N)^{-1}$ where N was the number of individuals with genotypes for the SNP within the analyzed breed. Among the variable SNPs, we further required that no more than 3 be contiguous. These thresholds were set to allow the detection of no more than one heterozygous individual for each variable SNP within a selective sweep region to allow for genotyping errors and for new variation to have been created within each region by mutation. These conditions also allow for the possibility that the SNPs may not have been correctly ordered by the UMD3.1 assembly and variable contigs may have been erroneously included within scaffolds containing a selective sweep.

### Annotation and functional analysis

Annotation of the genes present within all putative selective sweep regions was performed using the UCSC Genome Browser [71] and NCBI Gene database. Genes for which annotations were retrieved included any genes that were fully or partially contained within each region. Phenotypes known to be affected by variation in these genes were determined from a search of the literature and were assessed for their likely causality for each sweep. Functional analyses were performed for the sweeps detected within each breed using the functional annotation and clustering tools in the Database for Annotation, Visualization and Integrated Discovery (DAVID) [72].

## List of Abbreviations

Abbreviations are included in the Supplementary Information or are defined in text.

## Competing interests

The authors declare that they have no competing interests.

## Authors' contributions

JFT, JED and HRR designed the experiment. HRR, RDS and JFT analyzed data. SDM, MMR and RDS extracted DNA. MMR and SDM prepared samples for genotyping, SDM ran the Illumina assay, and RDS genotyped samples and managed the genotype database. HRR and JFT wrote the manuscript. All authors read and approved the final manuscript.

## Acknowledgements

We gratefully acknowledge the provision of semen samples from breeders and semen distributors. This project was supported by the University of Missouri, National Research Initiative grants number 2008-35205-04687 and 2008-35205-18864 from the USDA Cooperative State Research, Education and Extension Service and National Research Initiative grant number 2009-65205-05635 from the USDA National Institute of Food and Agriculture.

# References

1. Ross-Ibarra J, Morrell PL, Gaut BS: **Plant domestication, a unique opportunity to identify the genetic basis of adaptation.** *PNAS* 2007, **104:**8641-8648.
2. Loftus RT, MacHugh DE, Bradley DG, Sharp PM, Cunningham P: **Evidence for two independent domestications of cattle.** *PNAS* 1994, **91:**2757-2761.
3. Decker JE, Pires JC, Conant GC, McKay SD, Heaton MP, Chen K, Cooper A, Vilkki J, Seabury CM, Caetano AR, et al: **Resolving the evolution of extant and extinct ruminants with high-throughput phylogenomics.** *PNAS* 2009, **106:**18644-18649.
4. Brenneman RA, Davis SK, Sanders JO, Burns BM, Wheeler TC, Turner JW, Taylor JF: **The Polled locus maps to BTA1 in a *Bos indicus* x *Bos taurus* cross.** *J Hered* 1996, **87:**156-161.
5. Drögemüller C, Wöhlke A, Mömke S, Distl O: **Fine mapping of the polled locus to a 1-Mb region on bovine chromosome 1q12.** *Mamm Genome* 2005, **16:**613-620.
6. Medugorac I, Seichter D, Graf A, Russ I, Blum H, Göpel KH, Rothammer S, Förster M, Krebs S: **Bovine polledness – an autosomal dominant trait with allelic heterogeneity.** *PLoS One* 2012, **7:**e39477.
7. Klungland H, D I Vage LG-R, S Adalsteinsson and S Lien: **The role of melanocyte-stimulating hormone (MSH) receptor in bovine coat color determination.** *Mamm Genome* 1995, **6:**636-639.
8. Grosz MD, MacNeil MD: **The "spotted" locus maps to bovine chromosome 6 in a Hereford-cross population.** *J Hered* 1999, **90:**233-236.
9. Smith JM, Haigh J: **The hitch-hiking effect of a favourable gene.** *Genet Res* 1974, **23:**23-25.
10. Nielsen R, Hellmann I, Hubisz M, Bustamante C, Clark AG: **Recent and ongoing selection in the human genome.** *Nat Rev Genet* 2007, **8:**857-868.
11. Nielsen R: **Molecular signatures of natural selection.** *Annu Rev Genet* 2005, **39:**197-218.
12. The Bovine HapMap Consortium: **Genome-wide survey of SNP variation uncovers the genetic structure of cattle breeds.** *Science* 2009, **324:**528-532.
13. Nielsen R, Williamson S, Kim Y, Hubisz MJ, Clark AG, Bustamante C: **Genomic scans for selective sweeps using SNP data.** *Genome Res* 2005, **15:**1566-1575.
14. Sabeti PC, Schaffner SF, Fry B, Lohmueller J, Varilly P, Shamovsky O, Palma A, Mikkelsen TS, Altshuler D, Lander ES: **Positive natural selection in the human lineage.** *Science* 2006, **312:**1614-1620.
15. Rubin C-J, Zody MC, Eriksson J, Meadows JRS, Sherwood E, Webster MT, Jiang L, Ingman M, Sharpe T, Ka S, et al: **Whole-genome resequencing reveals loci under selection during chicken domestication.** *Nature* 2010, **464:**587-591.
16. Pollinger JP, Bustamante CD, Fledel-Alon A, Schmutz S, Gray MM, Wayne RK: **Selective sweep mapping of genes with large phenotypic effects.** *Genome Res* 2005, **15:**1809-1819.
17. Sutter NB, Bustamante CD, Chase K, Gray MM, Zhao K, Zhu L, Padhukasahasram B, Karlins E, Davis S, Jones PG, et al: **A single IGF1 allele is a major determinant of small size in dogs.** *Science* 2007, **316:**112-115.

18. Quilez J, Short AD, Martinez V, Kennedy LJ, Ollier W, Sanchez A, Altet L, Francino O: **A selective sweep of >8 Mb on chromosome 26 in the Boxer genome.** *BMC Genomics* 2011, **12:**12.
19. Makvandi-Nejad S, Hoffman GE, Allen JJ, Chu E, Gu E, Chandler AM, Loredo AI, Bellone RR, Mezey JG, Brooks SA, Sutter NB: **Four loci explain 83% of size variation in the horse.** *PLoS One* 2012, **7:**e39929.
20. Van Laere A-S, Nguyen M, Braunschweig M, Nezer C, Collette C, Moreau L, Archibald AL, Haley CS, Buys N, Tally M, et al: **A regulatory mutation in IGF2 causes a major QTL effect on muscle growth in the pig.** *Nature* 2003, **425:**832-836.
21. Andersson L: **How selective sweeps in domestic animals provide new insight into biological mechanisms.** *J Intern Med* 2012, **271:**1-14.
22. Decker JE, Vasco DA, McKay SD, McClure MC, Rolf MM, Kim JW, Northcutt SL, Bauck S, Woodward BW, Schnabel RD, Taylor JF: **A novel analytical method, birth date selection mapping, detects response of the Angus (*Bos taurus*) genome to artificial selection on complex traits.** *BMC Genomics,* under review.
23. Georges M, Drinkwater R, King T, Mishra A, Moore SS, Nielsen D, Sargeant LS, Sorensen A, Steele MR, Zhao X, et al: **Microsatellite mapping of a gene affecting horn development in *Bos taurus*.** *Nat Genet* 1993, **4:**206-210.
24. Qu Z, Adelson DL: **Bovine ncRNAs are abundant, primarily intergenic, conserved and associated with regulatory genes.** *PLoS One* 2012, **7:**e42638.
25. The Bovine Genome Sequencing and Analysis Consortium, Elsik CG, Tellam RL, Worley KC: **The genome sequence of taurine cattle: a window to ruminant biology and evolution.** *Science* 2009, **324:**522-528.
26. Margolis RL, McInnis MG, Rosenblatt A, Ross CA: **Trinucleotide repeat expansion and neuropsychiatric disease.** *Arch Gen Psychiatry* 1999, **56:**1019-1031.
27. Castermans D, Wilquet V, Parthoens E, Huysmans C, Steyaert J, Swinnen L, Fryns JP, Van de Ven W, Devriendt K: **The neurobeachin gene is disrupted by a translocation in a patient with idiopathic autism.** *J Med Genet* 2003, **40:**352-356.
28. Castermans D, Volders K, Crepel A, Backx L, De Vos R, Freson K, Meulemans S, Vermeesch JR, Schrander-Stumpel CT, De Rijk P, Del-Favero J, Van Geet C, Van De Ven WJ, Steyaert JG, Devriendt K, Creemers JW: **SCAMP5, NBEA and AMISYN: three candidate genes for autism involved in secretion of large dense-core vesicles.** *Hum Mol Gen* 2010, **19:**1368-1378.
29. Liu Z, Chang GQ, Leibowitz SF: **Diacylglycerol kinase Z in hypothalamus interacts with long form leptin receptor. Relation to dietary fat and body weight regulation.** *J Biol Chem* 2001, **276:**5900-5907.
30. Candille SI, Kaelin CB, Cattanach BM, Yu B, Thompson DA, Nix MA, Kerns JA, Schmutz SM, Millhauser GL, Barsh GS: **A β-defensin mutation causes black coat color in domestic dogs.** *Science* 2007, **318:**1418-1423.
31. Zhang S, Xu L, Lee J, Xu T: **Drosophila atrophin homolog functions as a transcriptional corepressor in multiple developmental processes.** *Cell* 2002, **108:**45-56.
32. Zoltewicz JS, Stewart NJ, Leung R, Peterson AS: **Atrophin 2 recruits histone deacetylase and is required for the function of multiple signaling centers during mouse embryogenesis.** *Development* 2004, **131:**3-14.

33. Plaster N, Sonntag C, Schilling TF, Hammerschmidt M: **REREa/Atrophin-2 interacts with histone deacetylase and Fgf8 signaling to regulate multiple processes of zebrafish development.** *Dev Dynam* 2007, **236:**1891-1904.
34. Bauersachs S, Mitko K, Ulbrich SE, Blum H, Wolf E: **Transcriptome studies of bovine endometrium reveal molecular profiles characteristic for specific stages of estrous cycle and early pregnancy.** *Exp Clin Endocrinol Diabetes* 2008, **116:**371-384.
35. Kageyama R, Ohtsuka T, Kobayashi T: **The Hes gene family: repressors and oscillators that orchestrate embryogenesis.** *Development* 2007, **134:**1243-1251.
36. Hansen TR, Austin KJ, Johnson GA: **Transient ubiquitin cross-reactive protein gene expression in the bovine endometrium.** *Endocrinology* 1997, **138:**5079-5082.
37. Bebington C, Bell SC, Doherty FJ, Fazleabas AT, Fleming SD: **Localization of ubiquitin and ubiquitin cross-reactive protein in human and baboon endometrium and decidua during the menstrual cycle and early pregnancy.** *Biol Reprod* 1999, **60:**920-928.
38. Cheutin T, McNairn AJ, Jenuwein T, Gilbert DM, Singh PB, Misteli T: **Maintenance of stable heterochromatin domains by dynamic HP1 binding.** *Science* 2003, **299:**721-725.
39. Martin C, Beaujean N, Brochard V, Audouard C, Zink D, Debey P: **Genome restructuring in mouse embryos during reprogramming and early development.** *Dev Biol* 2006, **292:**317-332.
40. Ruddock-D'Cruz NT, Prashadkumar S, Wilson KJ, Heffernan C, Cooney MA, French AJ, Jans DA, Verma PJ, Holland MK: **Dynamic changes in localization of chromobox (CBX) family members during the maternal to embryonic transition.** *Mol Reprod Dev* 2008, **75:**477-488.
41. Birling M-C, Tait S, Hardy RJ, Brophy PJ: **A novel rat tetraspan protein in cells of the oligodendrocyte lineage.** *J Neurochem* 1999, **73:**2600-2608.
42. Terada N, Baracskay K, Kinter M, Melrose S, Brophy PJ, Boucheix C, Bjartmar C, Kidd G, Trapp BD: **The tetraspanin protein, CD9, is expressed by progenitor cells committed to oligodendrogenesis and is linked to β1 integrin, CD81, and Tspan-2.** *Glia* 2002, **40:**350-359.
43. Arch JRS, Newsholme EA: **Activities and some properties of 5'-nucleotidase, adenosine kinase and adenosine deaminase in tissues from vertebrates and invertebrates in relation to the control of the concentration and the physiological role of adenosine.** *Biochem J* 1978, **174:**965-977.
44. Sudhof TC: **Neuroligins and neurexins link synaptic function to cognitive disease.** *Nature* 2008, **455:**903-911.
45. Kim H-G, Kishikawa S, Higgins AW, Seong I-S, Donovan DJ, Shen Y, Lally E, Weiss LA, Najm J, Kutsche K, et al: **Disruption of neurexin 1 associated with autism spectrum disorder.** *Am J Hum Genet* 2008, **82:**199-207.
46. Rouault TA, Tong WH: **Iron–sulfur cluster biogenesis and human disease.** *Trends in Genetics* 2008, **24:**398-407.
47. Lettre G, Jackson AU, Gieger C, Schumacher FR, Berndt SI, Sanna S, Eyheramendy S, Voight BF, Butler JL, Guiducci C, et al: **Identification of ten loci associated with height highlights new biological pathways in human growth.** *Nat Genet* 2008, **40:**584-591.

48. Karim L, Takeda H, Lin L, Druet T, Arias JAC, Baurain D, Cambisano N, Davis SR, Farnir F, Grisart B, et al: **Variants modulating the expression of a chromosome domain encompassing PLAG1 influence bovine stature.** *Nat Genet* 2011, **43:**405-413.
49. Dib MG, Van Vleck LD, Spangler ML: **Genetic analysis of mature size in American Angus cattle.** *Nebraska Beef Cattle Report* 2010**:**29-30.
50. Andersson EA, Harder MN, Pilgaard K, Pisinger C, Stančáková A, Kuusisto J, Grarup N, Færch K, Poulsen P, Witte DR, et al: **The birth weight lowering C-allele of rs900400 near *LEKR1* and *CCNL1* associates with elevated insulin release following an oral glucose challenge.** *PLoS One* 2011, **6:**e27096.
51. Weiss LA, Shen Y, Korn JM, Arking DE, Miller DT, Fossdal R, Saemundsen E, Stefansson H, Ferreira MAR, Green T, et al: **Association between microdeletion and microduplication at 16p11.2 and Autism.** *N Engl J Med* 2008, **358:**667-675.
52. Marshall CR, Noor A, Vincent JB, Lionel AC, Feuk L, Skaug J, Shago M, Moessner R, Pinto D, Ren Y, et al: **Structural variation of chromosomes in autism spectrum disorder.** *Am J Hum Gen* 2008, **82:**477-488.
53. Kumar RA, KaraMohamed S, Sudi J, Conrad DF, Brune C, Badner JA, Gilliam TC, Nowak NJ, Cook EH, Dobyns WB, Christian SL: **Recurrent 16p11.2 microdeletions in autism.** *Hum Mol Genet* 2008, **17:**628-638.
54. Chen DT, Jiang X, Akula N, Shugart YY, Wendland JR, Steele CJM, Kassem L, Park JH, Chatterjee N, Jamain S, et al: **Genome-wide association study meta-analysis of European and Asian-ancestry samples identifies three novel loci associated with bipolar disorder.** *Mol Psychiatry* 2011.
55. Burgess HA, Martinez S, Reiner O: **KIAA0369, doublecortin-like kinase, is expressed during brain development.** *J Neurosci Res* 1999, **58:**567-575.
56. Short B, Haas A, Barr FA: **Golgins and GTPases, giving identity and structure to the Golgi apparatus.** *Biochim Biophys Acta* 2005, **1744:**383-395.
57. Vjugina U, Zhu X, Oh E, Bracero NJ, Evans JP: **Reduction of mouse egg surface integrin alpha9 subunit (ITGA9) reduces the egg's ability to support sperm-egg binding and fusion.** *Biol Reprod* 2009, **80:**833-841.
58. Boyko AR, Quignon P, Li L, Schoenebeck JJ, Degenhardt JD, Lohmueller KE, Zhao K, Brisbin A, Parker HG, vonHoldt BM, et al: **A simple genetic architecture underlies morphological variation in dogs.** *PLoS Biol* 2010, **8:**e1000451.
59. Vaysse A, Ratnakumar A, Derrien T, Axelsson E, Pielberg GR, Sigurdsson S, Fall T, Seppälä EH, Hansen MST, Lawley CT, et al: **Identification of genomic regions associated with phenotypic variation between dog breeds using selection mapping.** *PLoS Genet* 2011, **7:**e1002316.
60. Sanders JO: **History and development of zebu cattle in the United States.** *J Anim Sci* 1980, **50:**1188-1200.
61. Campbell EMG, Nonneman D, Rohrer GA: **Fine mapping a quantitative trait locus affecting ovulation rate in swine on chromosome 8.** *J Anim Sci* 2003, **81:**1706-1714.
62. Chatterjee TK, Fisher RA: **Novel alternative splicing and nuclear localization of human RGS12 gene products.** *J Biol Chem* 2000, **275:**29660-29671.
63. Groenen MAM, Archibald AL, Uenishi H, Tuggle CK, Takeuchi Y, Rothschild MF, Rogel-Gaillard C, Park C, Milan D, Megens H-J, et al: **Analyses of pig genomes provide insight into porcine demography and evolution.** *Nature* 2012, **491:**393-398.

64. Nei M, Niimura Y, Nozawa M: **The evolution of animal chemosensory receptor gene repertoires: roles of chance and necessity.** *Nat Rev Genet* 2008, **9:**951-963.
65. Gilad Y, Wiebe V, Przeworski M, Lancet D, Pääbo S: **Loss of olfactory receptor genes coincides with the acquisition of full trichromatic vision in primates.** *PLoS Biol* 2004, **2:**e5.
66. Zhang X, De la Cruz O, Pinto JM, Nicolae D, Firestein S, Gilad Y: **Characterizing the expression of the human olfactory receptor gene family using a novel DNA microarray.** *Genome Biol* 2007, **8:**R86.
67. Purfield DC, Berry DP, McParland S, Bradley DG: **Runs of homozygosity and population history in cattle.** *BMC Genetics* 2012, **13:**70.
68. Van Tassell CP, Smith TP, Matukumalli LK, Taylor JF, Schnabel RD, Lawley CT, Haudenschild CD, Moore SS, Warren WC, Sonstegard TS: **SNP discovery and allele frequency estimation by deep sequencing of reduced representation libraries.** *Nat Methods* 2008, **5:**247-252.
69. Matukumalli LK, Lawley CT, Schnabel RD, Taylor JF, Allan MF, Heaton MP, O'Connell J, Moore SS, Smith TPL, Sonstegard TS, Tassell CPV: **Development and characterization of a high density SNP genotyping assay for cattle.** *PLoS One* 2009, **4:**e5350.
70. Rincon G, Weber KL, Van Eenennaam AL, Golden BL, Medrano JF: **Hot topic: Performance of bovine high-density genotyping platforms in Holsteins and Jerseys.** *J Dairy Sci* 2011, **94:**6116-6121.
71. Kent W, Sugnet C, Furey T, Roskin K, Pringle T, Zahler A, Haussler D: **The human genome browser at UCSC.** *Genome Res* 2002, **12:**996-1006.
72. Dennis Jr. G, Sherman BT, Hosack DA, Yang J, Baseler MW, Lane HC, Lempicki RA: **DAVID: Database for annotation, visualization, and integrated discovery.** *Genome Biol* 2003, **4:**P3.

**Figure 1. Selective sweep regions discovered in the analysis of the BovineSNP50 data that were predicted to be common to two or more breeds.**
Regions identified as harboring commonly selected haplotypes are indicated by the near-null MAF values and are indicated by black boxes.

**Figure 2. Selective sweep surrounding the *polled* locus in Angus cattle**
The selective sweep region on BTA1 in Angus is from ~1.7 cM to 2.0 cM and contains the *polled* locus. The locus is contained within an extended region of reduced heterozygosity relative to the MAF at down- and up-stream SNP.

**Table 1. Summary for genotyped individuals**

| Breed[1] | Origin | Primary Historical Use | Contiguous BovineSNP50 Loci[2] | Number Monomorphic SNP50 Loci | Number BovineSNP50 Individuals | Number AFFXB1P Individuals |
|---|---|---|---|---|---|---|
| Angus | Scotland | Beef | 6 | 8,443 | 2,918 | 23 |
| Braunvieh | Switzerland | Beef | 5 | 7,405 | 142 | |
| Charolais | France | Beef | 5 | 5,884 | 44 | |
| Hanwoo | Korea | Beef, Draft | 5 | 8,353 | 48 | 11 |
| Hereford | UK | Beef | 6 | 8,154 | 812 | |
| Limousin | France | Beef | 6 | 8,430 | 261 | |
| Salers | France | Beef | 6 | 8,409 | 72 | |
| Shorthorn | UK | Beef, Dairy | 6 | 8,558 | 108 | |
| Simmental | Switzerland | Beef, Dairy | 6 | 6,599 | 123 | 6 |
| Brown Swiss | Switzerland | Dairy | 9 | 12,553 | 74 | |
| Finnish Ayrshire | Scotland | Dairy | 5 | 6,185 | 599 | |
| Holstein | Netherlands | Dairy | 6 | 8,587 | 995 | |
| Jersey | Jersey | Dairy | 8 | 12,547 | 78 | |
| Wagyu | Japan | Beef, Draft | | | | 10 |
| Brahman | USA | Beef | 10 | 13,006 | 99 | 8 |
| **Total** | | | | | **6373** | **58** |

[1]Brahman are indicine cattle developed in the US as a composite of Nelore, Gir, Guzerat and Indu Brazil cattle originating in India but imported primarily from Brazil. The remaining breeds are taurine.
[2]Number of contiguous loci spanning at least 200 kb and with a minor allele frequency ≤ 0.01 required to declare a selective sweep region in each breed.

**Table 2. Putative selective sweep regions identified by analysis of BovineSNP50 genotypes.**

| Breed | BTA | UMD3.1 Coordinates (bp) | # SNP | Size (bp) |
|---|---|---|---|---|
| Angus[1] | 1 | 1,712,261-2,013,659 | 11 | 301,398 |
| Limousin | 2 | 5,974,885-6,344,425 | 7 | 369,540 |
| Brown Swiss | 2 | 73,436,684-73,978,469 | 10 | 541,785 |
| Hanwoo[1] | 3 | 19,860,064-20,175,158 | 5 | 315,094 |
| Hanwoo[1] | 3 | 112,997,892-113,219,287 | 5 | 221,395 |
| Brown Swiss | 4 | 61,098,696-61,376,132 | 9 | 277,436 |
| Hereford | 6 | 70,655,812-70,865,694 | 6 | 209,882 |
| Brown Swiss | 6[2] | 75,830,633-76,696,893 | 9 | 866,260 |
| Salers | | 75,996,320-76,696,893 | 8 | 700,573 |
| Jersey | 6 | 105,390,830-105,730,372 | 8 | 339,542 |
| Limousin | 7 | 40,250,259-40,485,825 | 7 | 235,566 |
| Jersey | 7 | 45,439,468-45,828,427 | 10 | 388,959 |
| Holstein | 7 | 72,908,532-73,126,315 | 7 | 217,783 |
| Brown Swiss | 11 | 25,577,969-25,941,999 | 10 | 364,030 |
| Jersey | 12 | 1,113,009-1,436,093 | 8 | 323,084 |
| Angus[1], Salers, Shorthorn, Simmental[1] | 12 | 25,878,820-26,236,394 | 7 | 357,574 |
| Brahman[1] | 12 | 36,287,533-36,989,957 | 14 | 702,424 |
| Holstein | 13 | 15,456,721-15,683,571 | 6 | 226,850 |
| Braunvieh | 14 | 28,674,493-28,960,475 | 5 | 285,982 |
| Brown Swiss | 14 | 42,739,573-43,140,076 | 9 | 400,503 |
| Braunvieh | 16 | 39,714,165-39,898,049 | 3 | 389,103 |
| Salers | 16 | 43,200,419-43,618,684 | 9 | 418,265 |
| Jersey | 16 | 45,376,614-45,874,144 | 8 | 497,530 |
| Angus[1], Holstein, Limousin, Simmental[1] | | 45,425,579-45,874,144 | 7 | 448,565 |
| Hereford | | 45,464,423-45,874,144 | 6 | 409,721 |
| Hanwoo[1] | 16 | 52,535,473-52,742,523 | 4 | 207,050 |
| Hanwoo[1] | 18 | 14,526,709-14,847,049 | 5 | 320,340 |
| Limousin | 19 | 37,301,353-37,520,214 | 6 | 218,861 |
| Brahman[1] | 19 | 43,376,032-43,835,219 | 10 | 459,187 |
| Angus[1], Holstein, Shorthorn | 24 | 63,576-320,143 | 6 | 256,567 |
| Braunvieh, Charolais, Finnish Ayrshire, Hanwoo[1] | | 105,589-320,143 | 5 | 214,554 |
| Angus[1], Holstein, Salers, Shorthorn | 27 | 25,075,449-25,295,935 | 6 | 220,486 |

[1]These breeds were also genotyped with the AFFXB1P assay and allow the potential validation of these putative selective sweep regions.

[2]Call rates for some SNP within this region did not exceed the 85% threshold that was applied to the remaining data.

**Table 3. Genomic regions predicted to harbor selective sweeps using BovineSNP50 data and validated by AFFXB1P data.**

| BTA | Breed | UMD3.1 Coordinates (bp) | # Fixed SNP | # Variable SNP | Size (bp) | Annotations |
|---|---|---|---|---|---|---|
| 1 | Angus | 1,673,108-2,024,737 | 313 | 8 | 351,629 | Horn-polled[6] |
| 13 | Wagyu | 15,493,906-15,739,251 | 301 | 4 | 245,345 | *DGKZ* |
| 16 | Angus | 45,386,065-45,652,672 | 238 | 10 | 266,607 | *ENO1*, *RERE* |
| | Simmental | 45,386,065-45,677,279 | 280 | 6 | 291,214 | |
| 16 | Simmental | 52,629,624-52,857,759 | 30 | 0 | 228,135 | *C16H1orf159*, *RNF223*, *F1N376*, *AGRN*, *ISG15*, |
| | Wagyu | 52,629,624-52,857,759 | 29 | 1 | 228,135 | *HES4*, *C16H1orf170*, *PLEKHN1*, *KLHL17*, *NOC2L*, *SAMD11* |
| 18 | Angus | 14,725,181-14,973,411 | 102 | 5 | 248,230 | *TCF25*, *SPIRE2*, *MC1R*, *TUBB3*, *MIR220D*, |
| | Simmental | 14,725,181-14,973,411 | 104 | 3 | 248,230 | *DEF8*, *CENPBD1*, *LOC532875*, *DBNDD1*, *GAS8*, *LOC100296324*, *LOC100336472*, *SHCBP1* |

**Table 4. Putative breed-specific selective sweeps identified using the AFFXB1P data**

| Breed | BTA | UMD3.1 Coordinates (bp) | Size (bp) | # Total SNPs | # Fixed SNP | # Variable SNP |
|---|---|---|---|---|---|---|
| Angus | 1 | 1,673,108-2,024,737 | 351,629 | 235 | 231 | 4 |
| | 1 | 111,659,758-111,900,241 | 240,483 | 231 | 221 | 10 |
| | 3 | 28,196,188-28,405,853 | 209,665 | 207 | 199 | 8 |
| | 6 | 6,308,164-6,696,861 | 388,697 | 74 | 72 | 2 |
| | 8 | 70,780,263-71,227,757 | 447,494 | 24 | 24 | 0 |
| | 8 | 73,538,014-73,760,303 | 222,289 | 288 | 282 | 6 |
| | 11 | 99,717,097-99,929,710 | 212,613 | 63 | 60 | 3 |
| | 14 | 3,550,689-3,885,375 | 334,686 | 21 | 21 | 0 |
| | 14 | 24,631,146-25,173,007 | 541,861 | 387 | 376 | 11 |
| | 18 | 200,903-406,270 | 205,367 | 117 | 114 | 3 |
| | 20 | 70,827,025-71,040,113 | 213,088 | 30 | 29 | 1 |
| | 21 | 1,186,567-1,489,860 | 303,293 | 26 | 25 | 1 |
| | 21 | 3,091,822-3,417,143 | 325,321 | 308 | 301 | 7 |
| | 25 | 39,262,249-39,468,067 | 205,818 | 21 | 20 | 1 |
| | 27 | 36,633,324-36,896,534 | 263,210 | 135 | 130 | 5 |
| | 29 | 49,557,166-49,922,377 | 365,211 | 70 | 69 | 1 |
| Brahman | 5 | 48,679,627-48,903,409 | 223,782 | 249 | 237 | 12 |
| | 10 | 24,519,718-24,794,435 | 274,717 | 20 | 20 | 0 |
| | 22 | 10,701,509-10,963,520 | 262,011 | 88 | 86 | 2 |
| Hanwoo | 9 | 1,384,189-1,586,988 | 202,799 | 214 | 211 | 3 |
| | 22 | 60,772,158-61,015,491 | 243,333 | 24 | 23 | 1 |
| Simmental | 1 | 83,467,837-83,685,705 | 217,868 | 124 | 118 | 6 |
| | 2 | 36,576,839-36,828,533 | 251,694 | 216 | 206 | 10 |
| | 2 | 119,873,146-120,084,764 | 211,618 | 158 | 151 | 7 |
| | 2 | 121,398,894-121,698,563 | 299,669 | 196 | 188 | 8 |
| | 5 | 68,631,784-68,857,802 | 226,018 | 276 | 267 | 9 |
| | 7 | 21,101,050-21,348,345 | 247,295 | 65 | 63 | 2 |
| | 7 | 52,426,108-52,673,522 | 247,414 | 40 | 38 | 2 |
| | 8 | 70,246,437-70,540,424 | 293,987 | 163 | 158 | 5 |
| | 10 | 59,147,189-59,382,774 | 235,585 | 190 | 182 | 8 |
| | 10 | 72,811,087-73,041,092 | 230,005 | 217 | 210 | 7 |
| | 13 | 12,236,039-12,524,368 | 288,329 | 382 | 369 | 13 |
| | 13 | 66,468,866-66,708,674 | 239,808 | 153 | 146 | 7 |
| | 15 | 82,007,368-82,208,624 | 201,256 | 32 | 32 | 0 |
| | 16 | 42,555,966-42,788,613 | 232,647 | 46 | 44 | 2 |
| | 16 | 43,767,572-44,053,904 | 286,332 | 186 | 181 | 5 |
| | 17 | 13,291,922-13,492,010 | 200,088 | 141 | 136 | 5 |
| | 19 | 34,588,859-35,435,279 | 846,420 | 59 | 57 | 2 |
| | 22 | 2,788,408-3,001,597 | 213,189 | 216 | 214 | 2 |
| | 25 | 26,433,626-26,744,488 | 310,862 | 126 | 120 | 6 |
| | 25 | 42,202,212-42,775,697 | 573,485 | 21 | 21 | 0 |
| | 28 | 30,480,640-30,711,987 | 231,347 | 254 | 243 | 11 |
| | 29 | 39,045,837-39,396,761 | 350,924 | 93 | 91 | 2 |
| | 29 | 49,052,295-49,273,683 | 221,388 | 21 | 20 | 1 |
| Wagyu | 1 | 197,944-574,931 | 376,987 | 321 | 313 | 8 |
| | 1 | 12,544,028-12,840,164 | 296,136 | 335 | 325 | 10 |
| | 2 | 71,299,739-71,508,079 | 208,340 | 177 | 169 | 8 |
| | 2 | 136,434,039-136,676,537 | 242,498 | 22 | 21 | 1 |
| | 3 | 12,027,611-12,227,923 | 200,312 | 65 | 64 | 1 |
| | 3 | 113,841,208-114,151,444 | 310,236 | 189 | 181 | 8 |
| | 3 | 117,817,862-118,017,993 | 200,131 | 114 | 110 | 4 |
| | 4 | 117,846,325-118,095,397 | 249,072 | 21 | 20 | 1 |
| | 4 | 119,437,108-119,832,270 | 395,162 | 26 | 25 | 1 |
| | 11 | 33,020,266-33,264,604 | 244,338 | 307 | 293 | 14 |
| | 11 | 59,513,020-59,783,144 | 270,124 | 282 | 269 | 13 |
| | 11 | 103,333,722-103,590,033 | 256,311 | 33 | 32 | 1 |
| | 11 | 105,596,531-105,828,115 | 231,584 | 23 | 22 | 1 |
| | 12 | 23,283-377,340 | 354,057 | 137 | 137 | 0 |
| | 13 | 15,493,906-15,739,251 | 245,345 | 305 | 301 | 4 |
| | 13 | 58,016,490-58,335,951 | 319,461 | 114 | 110 | 4 |
| | 16 | 51,275,719-51,519,519 | 243,800 | 39 | 38 | 1 |
| | 19 | 53,829,568-54,065,288 | 235,720 | 25 | 24 | 1 |
| | 22 | 49,088,027-49,456,146 | 368,119 | 102 | 98 | 4 |
| | 24 | 36,584,702-36,795,576 | 210,874 | 224 | 218 | 6 |
| | 27 | 17,911,399-18,114,646 | 203,247 | 304 | 289 | 15 |

**Table 5. Putative selective sweep regions detected in at least two breeds using AFFXB1P data**

| Breeds | BTA | UMD3.1 Coordinates (bp) | Size (bp) | # SNPs Total | # SNPs Fixed | # SNPs Variable |
|---|---|---|---|---|---|---|
| Angus | 6 | 5,639,799-5,993,214 | 353,415 | 95 | 92 | 3 |
| Simmental, Wagyu, Hanwoo | | 5,639,799-5,993,214 | 353,415 | 95 | 94 | 1 |
| Angus, Simmental, Wagyu | 6 | 106,844,116-107,308,280 | 464,164 | 21 | 20 | 1 |
| Simmental | 7 | 4,344,221-4,677,474 | 333,253 | 22 | 21 | 1 |
| Hanwoo | | 4,443,937-4,838,781 | 394,844 | 90 | 88 | 2 |
| Angus | 7 | 51,000,141-51,430,242 | 430,101 | 406 | 391 | 15 |
| Simmental | | 51,144,109-51,788,442 | 644,333 | 477 | 469 | 8 |
| Wagyu | 14 | 2,098,182-2,588,143 | 489,961 | 29 | 29 | 0 |
| Hanwoo | | 2,097,493-2,603,935 | 506,442 | 31 | 30 | 1 |
| Angus | 16 | 44,463,682-44,881,229 | 417,547 | 213 | 203 | 10 |
| Simmental | | 44,693,447-45,207,060 | 513,613 | 162 | 158 | 4 |
| Angus | 16 | 45,386,065-45,652,672 | 266,607 | 248 | 238 | 10 |
| Simmental | | 45,386,065-45,677,279 | 291,214 | 286 | 280 | 6 |
| Simmental | 16 | 49,400,423-49,679,673 | 279,250 | 22 | 21 | 1 |
| Wagyu | | 49,400,423-49,661,831 | 261,408 | 20 | 20 | 0 |
| Simmental | 16 | 52,629,624-52,857,759 | 228,135 | 30 | 30 | 0 |
| Wagyu | | 52,629,624-52,857,759 | 228,135 | 30 | 29 | 1 |
| Wagyu | 17 | 69,501,174-69,703,717 | 202,543 | 116 | 112 | 4 |
| Wagyu | | 69,953,291-70,183,461 | 230,170 | 145 | 140 | 5 |
| Hanwoo | | 69,705,850-70,283,199 | 577,349 | 421 | 402 | 19 |
| Simmental | 17 | 73,619,837-73,975,539 | 355,702 | 24 | 24 | 0 |
| Wagyu | | 73,761,834-74,325,363 | 563,529 | 57 | 55 | 2 |
| Angus | 18 | 14,725,181-14,973,411 | 248,230 | 107 | 102 | 5 |
| Simmental | | 14,725,181-14,973,411 | 248,230 | 107 | 104 | 3 |
| Angus | 18 | 53,342,619-53,596,733 | 254,114 | 53 | 51 | 2 |
| Simmental | | 53,342,619-53,573,919 | 231,300 | 46 | 44 | 2 |
| Angus | 19 | 57,106,999-57,377,040 | 270,041 | 26 | 25 | 1 |
| Simmental | | 57,065,941-57,377,040 | 311,099 | 28 | 28 | 0 |
| Angus | 20 | 71,679,859-72,012,001 | 332,142 | 37 | 36 | 1 |
| Simmental | | 71,780,338-72,012,001 | 231,663 | 29 | 28 | 1 |
| Angus | 21 | 2,134-742,281 | 740,147 | 422 | 408 | 14 |
| Simmental | | 2,134-326,342 | 324,208 | 187 | 179 | 8 |
| Angus, Simmental | 21 | 1,727,412-2,142,823 | 415,411 | 30 | 29 | 1 |
| Angus | 21 | 70,948,810-71,284,393 | 335,583 | 23 | 22 | 1 |
| Hanwoo | | 71,112,766-71,575,370 | 462,604 | 32 | 31 | 1 |
| Angus | 26 | 21,432-718,976 | 697,544 | 242 | 234 | 8 |
| Simmental | | 21,432-723,266 | 701,834 | 250 | 244 | 6 |
| Wagyu | | 21,432-225,284 | 203,852 | 30 | 29 | 1 |
| Wagyu | | 249,534-454,031 | 204,497 | 81 | 77 | 4 |
| Hanwoo | | 21,432-454,031 | 432,599 | 121 | 119 | 2 |
| Simmental | 29 | 1,156-558,130 | 556,974 | 155 | 153 | 2 |
| Wagyu | | 105,179-536,644 | 431,465 | 122 | 120 | 2 |
| Hanwoo | | 315,440-558,130 | 242,690 | 90 | 86 | 4 |

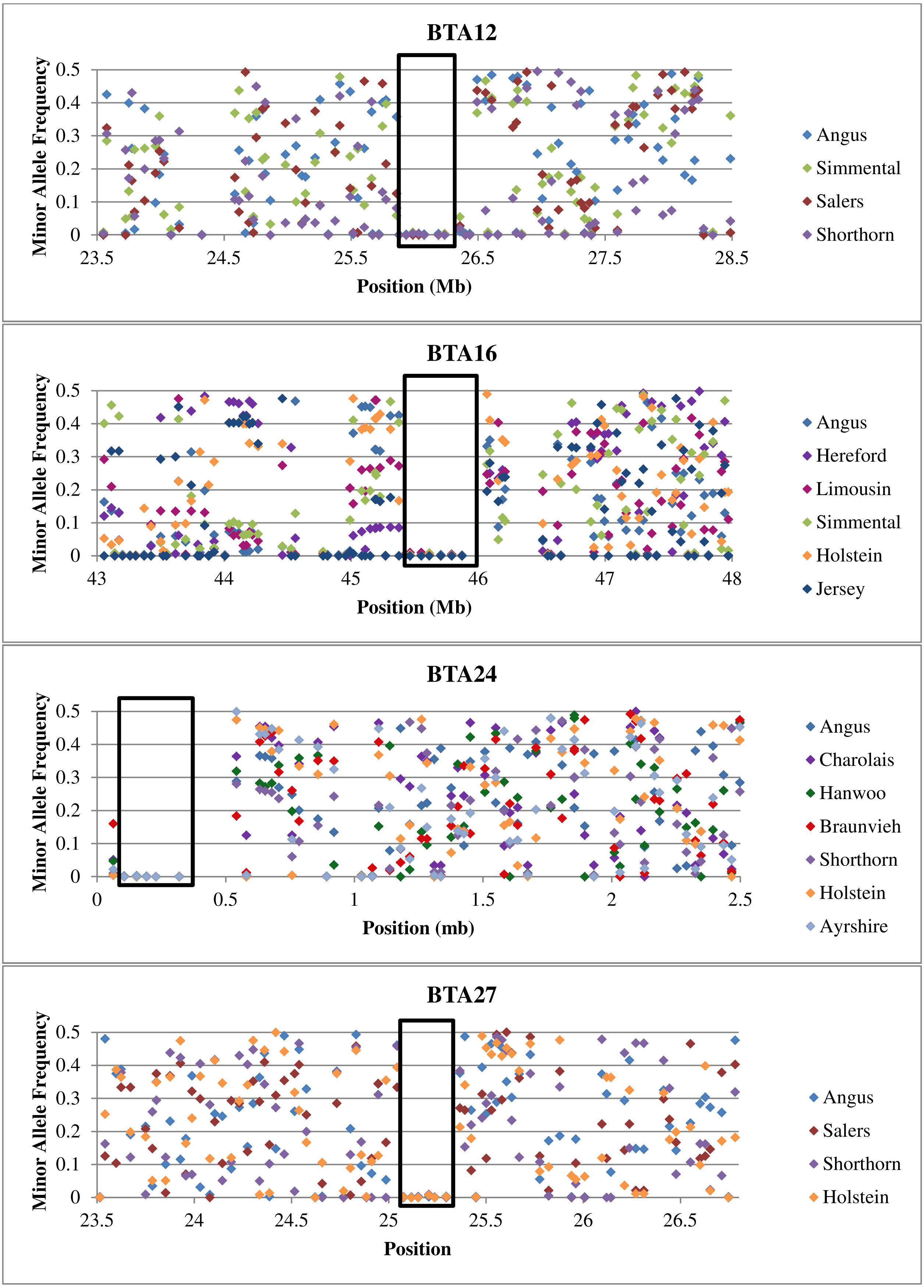

Figure 1

Figure 2

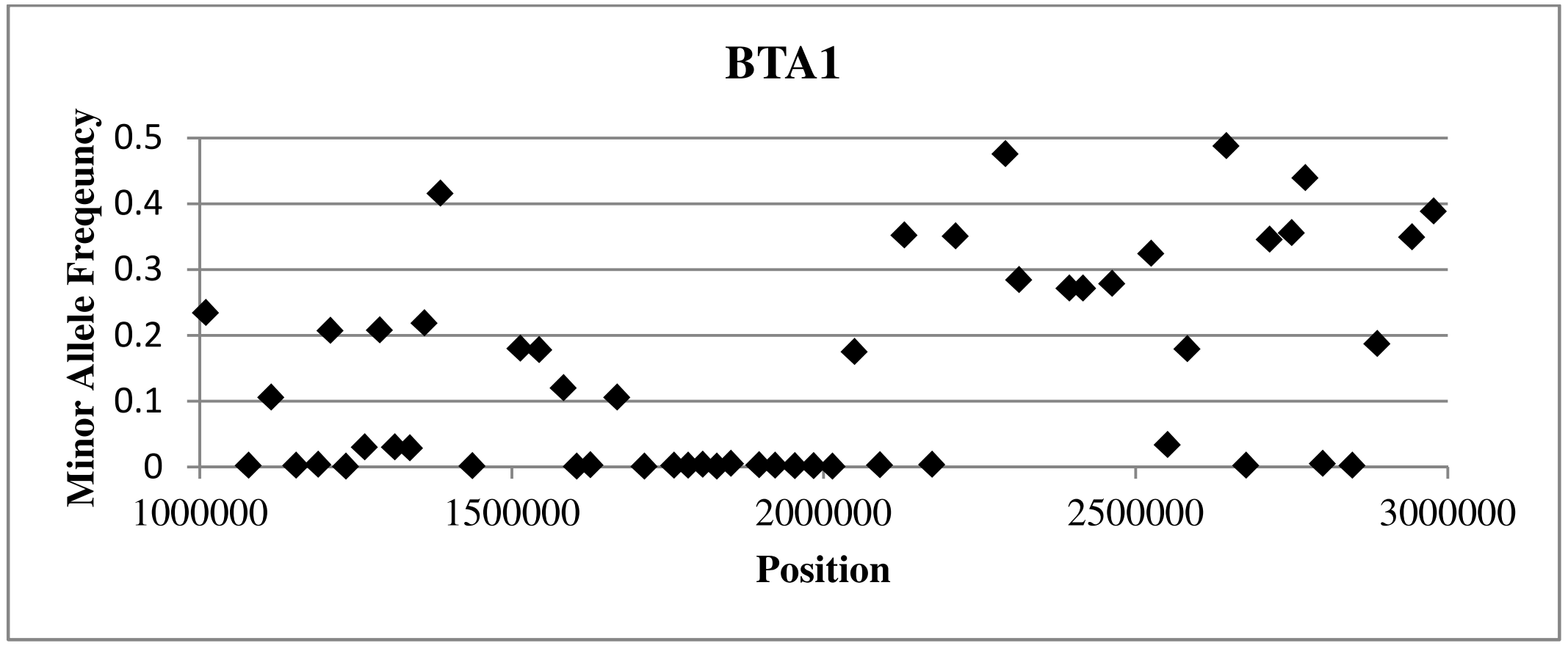